\documentclass[fleqn,usenatbib]{mnras}

\usepackage{newtxtext,newtxmath}
\usepackage[T1]{fontenc}

\DeclareRobustCommand{\VAN}[3]{#2}
\let\VANthebibliography\thebibliography
\def\thebibliography{\DeclareRobustCommand{\VAN}[3]{##3}\VANthebibliography}

\usepackage{graphicx}	% Including Figure files
\usepackage{amsmath}	% Advanced maths commands
\usepackage{caption}
\usepackage{xcolor}
\usepackage{subcaption}
\usepackage[normalem]{ulem} % For \sout{}

\newcommand{\soutPC}{\bgroup\markoverwith{\textcolor{cyan}{\rule[0.5ex]{2pt}{1pt}}}\ULon}

\newcommand{\soutdif}{\bgroup\markoverwith{\textcolor{magenta}{\rule[0.5ex]{2pt}{1pt}}}\ULon}
\newcommand\T{\rule{0pt}{2.6ex}}       % Top strut
\newcommand\B{\rule[-1.2ex]{0pt}{0pt}} % Bottom strut

\title[Proto-globular clusters at $z>4$]{Stellar cluster formation in a Milky Way-sized galaxy at $z>4$ -- I. The proto-globular cluster population and the imposter amongst us}

\author[F. van Donkelaar et al.] 
{Floor van Donkelaar,$^{1}$\thanks{floor.vandonkelaar@uzh.ch} Lucio Mayer,$^{1}$ Pedro R. Capelo,$^{1}$ Tomas Tamfal,$^{1}$ Thomas R. Quinn$^{2}$ \newauthor and Piero Madau$^{3}$\\
$^1$Center for Theoretical Astrophysics and Cosmology, Institute for Computational Science, University of Zurich,\\ Winterthurerstrasse 190, CH-8057 Z\"urich, Switzerland\\
$^2$Astronomy Department, University of Washington, Seattle, WA 98195, USA\\
$^3$Department of Astronomy and Astrophysics, University of California, 1156 High Street, Santa Cruz, CA 95064, USA}

\date{Accepted XXX. Received YYY; in original form ZZZ}

\pubyear{2023}

\begin{document}
\label{firstpage}
\pagerange{\pageref{firstpage}--\pageref{lastpage}}
\maketitle

\begin{abstract}
The formation history of globular clusters (GCs) at redshift $z > 4$ remains an unsolved problem. In this work, we use the cosmological, $N$-body hydrodynamical ``zoom-in'' simulation GigaEris to study the properties and formation environment of proto-GC candidates in the region surrounding the progenitor of a  Milky Way-sized galaxy. The simulation employs a modern implementation of smoothed-particle hydrodynamics, including metal-line cooling and metal and thermal diffusion. We define proto-GC candidate systems as gravitationally bound stellar systems with baryonic mass fraction $F_{\rm b} \geq 0.75$ and stellar velocity dispersion $\sigma_{\star} < 20$~km~s$^{-1}$. At $z=4.4$ we identify 9 systems which satisfy our criteria, all of which form between 10 kpc to 30 kpc from the centre of the main host. Their baryonic masses are in the range $10^5$--$10^7$~M$_{\sun}$. By the end of the simulation, they still have a relatively low stellar mass ($M_{\star} \sim 10^4$--$10^5$~M$_{\sun}$) and a metallicity ($-1.8 \lesssim {\rm [Fe/H]} \lesssim -0.8$) similar to the blue Galactic GCs. All of the identified systems except one appear to be associated with gas filaments accreting onto the main galaxy in the circum-galactic region, and formed at $z=5$--$4$. The exception is the oldest object, which appears to be a stripped compact dwarf galaxy that has interacted with the main halo between $z = 5.8$ and $z=5.2$ and has lost its entire dark matter content due to tidal mass loss.
\end{abstract}


%%%%%%%%%%%%%%%%%%%%%%%%%%%%%%%%%%%%%%%%%%%%%%%%%%

%%%%%%%%%%%%%%%%% BODY OF PAPER %%%%%%%%%%%%%%%%%%

\section{Introduction}

Although globular clusters (GCs) are very well studied in our local Galactic neighbourhood, their formation and evolution at redshift $z~>~4$ remains an unsolved problem. GCs are spheroidal, gravitationally bound collections of stars that are in general very old \citep[$\sim$13~Gyr; e.g.][]{Trenti:2015aa} structures with stellar masses $\sim$  $10^5$--$10^6$~M$_{\sun}$ and are observed to be metal poor \citep[$-2.5 \lesssim {\rm [Fe/H]} \lesssim 0.3$; see, e.g.][]{Harris:1991aa, Elmegreen:1997aa, Fall:2001aa, Brodie:2006aa, Mclaughlin:2008aa, Elmegreen:2010aa}. Furthermore, GCs are found in almost all galaxies with stellar mass $M_{\star} > 10^9$~M$_{\sun}$ in the local Universe. As they are amongst the oldest astrophysical objects in the Universe, GCs have witnessed most of the formation and evolution processes of galaxies, and can be used to probe them \citep{Brodie:2006aa}.

Since the environmental conditions that are required for and the nature of the process of formation of GCs are still unknown, a number of theories have been proposed to explain their formation. \citet{Peebles:1968aa}, for example, suggested that GCs formed soon after recombination in low-metallicity environments, with masses regulated by the cosmological \citet{Jeans1902} mass. Similarly, \cite{Peebles:1984aa} proposed that these GCs were formed in dark matter (DM) halos at high redshift and later merged into more massive halos. These theories, however, would not result in the colour bimodality -- the blue and red clusters -- observed  \citep[e.g.][]{Gebhardt:1999aa,Peng:2006aa}. Another theory that has been suggested is that GCs were formed by thermal instability in hot gaseous halos of young galaxies \citep[e.g.][]{Fall:1985aa, Kang:1990aa}. Accordingly, \cite{Ashman:1922aa} suggested a two-step formation channel for the formation of GCs that  would account for the bimodality. They stated that the blue and metal-poor GCs form via the scenario described by \citeauthor{Peebles:1968aa}, but that the red and metal-rich GCs are formed through star formation (SF) that is triggered during mergers, whereby disc galaxies merge to form elliptical galaxies. This model predicted multi-model colour distributions in elliptical galaxies \citep[see also][for GCs formed as a byproduct of active SF in galaxy discs]{Elmegreen:2010aa, Shapiro:2010aa, Kruijssen:2015aa}. \citet{gnedin_2009}, for example, showed that early mergers of smaller host galaxies create exclusively blue clusters, whereas subsequent mergers of progenitor galaxies with a range of different masses create both red and blue clusters. Thus, bimodality arises naturally as the result of a small number of late massive merger events \citep[see also][]{Li:2004aa, Kravtov:2005aa, Li:2017aa, Pfeffer:2018aa, keller:2020aa, Li:2022aa}.  \citet{Rosenblatt:1988aa} improved the \citeauthor{Peebles:1968aa} scenario by assuming that pre-galactic GCs formed within cold DM halos, but only those formed in high-$\sigma$ fluctuations would survive. Another popular mechanism is that GCs are formed by the collisions of DM haloes at high redshift \citep[][]{Beasley:2020aa, Madau:2020aa}, resulting in colliding substructures with stellar masses of $10^{2.6} \leq M_{\star}[{\rm M}_{\sun}] \leq 10^{8.6}$ with a median value of $10^{4.4}$~M$_{\sun}$. This proposed mechanism by \citet{Madau:2020aa} gives origin to ``naked'' GCs as colliding DM subhalos and their stars will pass through one another. 

Significantly, observations also suggest that GCs contain little to no DM \citep[e.g.][]{Conroy:2011aa, Ibata:2013aa}.  This is an important aspect that any formation theory of GCs needs to encompass. GCs could therefore have potentially formed as a result of gravitational instabilities driven by baryons without the need for DM, for example through supersonically induced gas objects \citep{Lake:2022aa}. Another possibility is that GCs were formed within the local potential minimum generated by a DM halo, but that these halos were later stripped by the tidal field of their host galaxies \citep[e.g.][]{Bromm:2002aa}. 

In order to test formation hypotheses of the GCs observed in the local Universe, detailed spectroscopic and astrometric observations of stars in the vicinity of GCs are required and the structural and kinematic properties of these stellar systems must be studied. For example, using kinematic information provided by Gaia \citep{gaia:2018aa}, \citet{Massari:2019aa} found that about 40 per cent of the Galactic GCs in the Milky Way likely formed in situ. These kind of observations have only become possible recently, as a result of on-going missions like Gaia. To directly discover the exact formation channel for GCs via observations in the early Universe, high-redshift observations are needed. Recent evidence shows that possible proto-GCs may already have been detected \citep[e.g.][]{Vanzella:2017aa, Elmegreen:2017aa, Kiku:2020aa}. These observations, however, do not yet provide the resolution and detail required to assess the implications of the different formation scenarios. Nevertheless, as we enter the era of the James Webb Space Telescope (JWST), the Extremely Large Telescopes (ELTs), the Atacama Large Millimitre/submillimetre Array (ALMA), and the Square Kilometre Array (SKA), the opportunity to observe these possible proto-GCs and determine their formation path may soon be on the horizon \citep[e.g.][]{Katz:2013aa,Hessels:2015aa, Renzini:2017aa, Pozzetti:2019aa, Calura:2021aa}.

Numerical simulations have attempted to explain the formation of GCs and test the  current theoretical models.  Over recent years, the resolution of cosmological simulations has enhanced greatly and the models describing the underlying physics have been significantly improved. This allows us to probe with much greater physical realism than before the early stages of galaxy formation, when the progenitors of present-day Milky Way-like galaxies have a baryonic and DM mass considerably smaller than at the present day. To study the formation of GCs, numerical investigations are conducted either in isolated \citep[e.g.][]{Nakasato:200aa, Bate2003} or cosmological settings \citep[e.g.][]{Kravtov:2005aa}. Several groups now also use a variety of simulation suites, ``zoom-in'' simulations, to study the formation of GCs \citep[e.g.][]{Rieder:2013, Ricotti:2016aa, Renaud:2017aa, Kim:2018aa, Ma:2020aa, Reina:2022aa}. In ``zoom-in'' simulations, a region of interest is simulated at high resolution, while the surroundings of that region is left at a lower resolution. The low-resolution surroundings provide a realistic tidal field for the high-resolution region of interest. This technique allows for single halos to be studied in great detail, without resorting to non-cosmological simulations of isolated halos. Using ``zoom-in'' simulations, we can better resolve the internal structures of galaxies. This allows us for example to capture the multi-phase interstellar medium, the SF, and the evolution of the GCs.

In this work, we revisit the challenge of the formation and the birth properties of (proto-) GCs using a ``zoom-in'' cosmological hydrodynamical simulation of unprecedented resolution, GigaEris \citep{Tamfal:2022aa}. The layout of the paper is the following: Section~\ref{sec:method}  briefly summarises the simulation setup and describes how we identified the halos. In Section~\ref{sec:results}, we present the simulation results, with a focus on the properties of the proto-GCs at $z = 4.4$. Furthermore, we have a brief look at the evolution of these objects. Finally, we discuss our results in Section~\ref{sec:disc} and conclude in Section~\ref{sec:conc}. 

\section{Methods}\label{sec:method}

\subsection{Simulation code \& Initial conditions}

This work uses the GigaEris cosmological, $N$-body hydrodynamical ``zoom-in'' simulation. A brief summary of the numerical recipe is provided below, whereas a more detailed description can be found in \citet{Tamfal:2022aa}.

The simulation was run using \textsc{ChaNGa} \citep[][]{Jetley:2008aa,Jetley:2010aa, Menon:2015aa}, an $N$-body smoothed-particle hydrodynamics (SPH) code \citep[][]{Wadsley:2017aa}. \citet{Tamfal:2022aa} follow a Galactic-scale halo identified at $z = 0$ in a periodic cube of side 90 cMpc. In the code, each individual stellar particle represents an entire stellar population following the initial mass function described in \citet{Kroupa:2001aa}, with an initial particle mass of $m_{\star} = 798$~M$_{\sun}$. Stars are formed stochastically using a simple gas density and temperature threshold criterion, with $n_{\rm SF} > 100$~atoms~cm$^{\text{-} 3}$ and $T_{\rm SF} < 3 \times 10^4$~K, and with an SF rate (SFR) given by

\begin{equation}
    \frac{\rm{d} \rho_{\star}}{\rm{d}t} = \epsilon_{\rm SF} \frac{\rho_{\rm gas}}{t_{\rm dyn}},
\end{equation}

\noindent with $\rho_{\star}$ denoting the stellar density, $\rho_{\rm gas}$ the gas density, $t_{\rm dyn}$ the local dynamical time, and $\epsilon_{\rm SF}$ the SF efficiency, which is set to $0.1$.  The code solves for the non-equilibrium abundances and cooling of H and He  species, assuming self-shielding \citep[see][]{Pontzen:2008aa} and a redshift-dependent ultraviolet radiation background \citep{Haardt:2012aa}, whereas the cooling from the fine structure lines of metals is calculated in photoionization equilibrium from the same radiation background (assuming no self-shielding, see \citealt{Capelo:2018aa} for a discussion), using tabulated rates from Cloudy \citep{Ferland:2010aa, Ferland:2013aa} and following the method described in \citet{Shen:2010aa, Shen:2013aa}. For all species, cooling is modelled down to 10~K.

Feedback from supernovae Type Ia is implemented by injecting energy and a fixed amount of mass and metals, independent of the progenitor mass, into the surroundings \citep[][]{Thielemann:1986, Stinson:2006aa}, whereas supernovae Type II (SNII) feedback is implemented following the delayed-cooling recipe of \citet{Stinson:2006aa}, each injecting metals and $\epsilon_{\rm SF} =10^{51}$ erg per event into the interstellar medium as thermal energy, according to the ‘blastwave model’ of \citet{Stinson:2006aa}. Stars with masses between 8 and 40~M$_{\sun}$ can explode as SNII, furthermore for each SNII event a given amount of oxygen and iron mass, dependent on the mass of the star, is injected into the surrounding gas \citep{Woosley:1995aa, raiteri:1996aa}. Stars with masses 1~$< m_{\star} <$~8~M$_{\sun}$ do not explode as supernovae but release part of their mass as stellar winds with the returned gas having the same metallicity of the low-mass stars. 

The initial conditions are implemented using the \textsc{MUSIC} code \citep[][]{Hahn:2011aa}, using 14 levels of refinement and the cosmological parameters $\Omega_{\rm m}$ = 0.3089, $\Omega_{\rm b}$ = 0.0486, $\Omega_{\Lambda}$ = 0.6911, $\sigma_8$ = 0.8159, $n_{\rm s}$ = 0.9667, and $H_0$ = 67.74~km~s$^{-1}$ Mpc$^{-1}$ \citep[see][]{Planck:2016aa}. The simulation reached $z=4.4$, when the particle numbers in the entire simulation box are $n_{\rm DM} = 5.7 \times 10^8$, $n_{\rm gas} = 5.2 \times 10^8$, and $n_{\star} = 4.4 \times 10^7$.

\subsubsection{The spatial resolution}

The gravitational softening of all particles is constant in physical coordinates ($\epsilon_{\rm c} = 0.043$~kpc) for redshifts smaller than $z = 10$ and otherwise evolves as $\epsilon = 11\epsilon_{\rm c}/(1+z)$. As a result, the gravitational softening at $z=4.44$ is 43 pc, which is larger than the half-mass radius of many of the known GCs \citep[e.g.][]{Madrid:2012aa, Piatti:2019aa,Vasiliev:2021aa}. Conversely, the SPH smoothing length, owing to the very high mass resolution of the GigaEris simulation, is around 13~pc in proto-GCs right after formation, hence about 3 times smaller than the gravitational softening. As it has been thoroughly studied and demonstrated in the literature on  SF simulations, when the smoothing length is appreciably smaller then the softening in a self-gravitating system, gravitational collapse is slowed down or even partially suppressed \citep[e.g.][]{Bate:1997aa}. Hence the radii measured in our simulations should be considered as upper limits to the actual radii of the proto-GCs. With smaller radii, proto-GCs will reach higher densities and undergo faster SF, thus building up faster a larger stellar mass. Compared to the results that we will show in the following sections, the match with observations could improve in more than one way with smaller softening. Nevertheless, aside from the smaller radii, since here we analyze the properties of proto-GCs soon after their formation, we expect results on global properties such as stellar mass, metallicity, and kinematics to vary only slightly.

\subsection{Halo finding}

In order to find the substructures that possibly could be proto-GCs, the \textsc{AMIGA Halo Finder} \citep[hereafter \textsc{AHF},][]{Gill:2004aa, Knollmann:2009aa} was applied to our simulation box with a minimum number of 64 baryonic particles per halo. To locate the possible proto-GCs, we only took substructures with 0 subhalos inside the identified halo. Hereafter the word halo refers to a bound object found by \textsc{AHF}. To confirm that the systems are bound, we have calculated the binding energies of the identified halos.

\section{Results} \label{sec:results}

\begin{figure}
\centering
\setlength\tabcolsep{2pt}%
\includegraphics[ trim={0cm 0cm 0cm 0cm}, clip, width=0.48\textwidth, keepaspectratio]{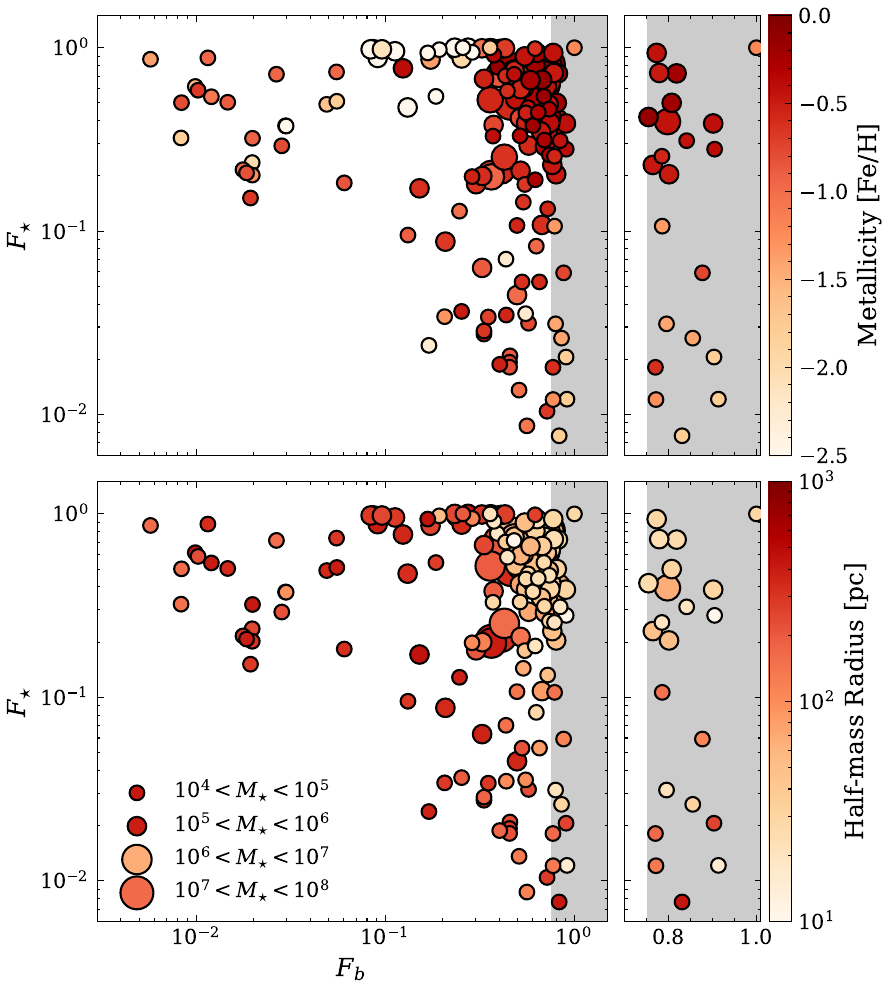}
\caption{Stellar mass fraction ($F_{\star}$) plotted against the baryonic mass fraction ($F_{\rm b}$) for all halos identified by \textsc{AHF} in the stellar mass range $10^4 < M_{\star}[{\rm M}_{\sun}] < 10^8$ at $z=4.4$ (left-hand panels) and for all these identified halos with $F_{\rm b} \geq 0.75$ (right-hand panels). The size of the markers indicates the stellar mass range, computed at half the virial radius. The gray-shaded area indicates the region where  $F_{\rm b} \geq 0.75$. From top to bottom, the colour bars represents the stellar metallicity and the half-mass radius, $R_{\rm m}$, of the halos.}
\label{fig:Fs_Fb}
\end{figure}

\subsection{Classifying the proto-globular cluster systems}\label{sec:class}

To investigate the properties, formation, and evolution of proto-GC candidates within our simulation, we extract all AHF identified substructures in the simulated box within the stellar mass range $10^4$--$10^8$~M$_{\sun}$. For each of the substructures, the baryonic mass fraction,

\begin{equation} \label{fb}
    F_{\rm b} = \frac{(M_{\star}+M_{\rm gas})}{M_{\rm total}},
\end{equation}

\noindent and stellar mass fraction,

\begin{equation} \label{fs}
    F_{\star} = \frac{M_{\star}}{(M_{\star}+M_{\rm gas})},
\end{equation}

\noindent are calculated, where $M_{\rm total}$ is the sum of the baryonic and DM masses, $M_{\star}$ is the stellar mass, and $M_{\rm gas}$ is the gas mass, all computed at half the virial radius. In Figure~\ref{fig:Fs_Fb}, the stellar mass fraction of the bound extracted objects (with stellar mass $10^4 \leq M_{\star}[{\rm M}_{\sun}] \leq 10^8$) is plotted against the baryonic mass fraction at $z=4.4$. We select all halos with a baryonic mass fraction  $F_{\rm b} \geq 0.75$ (this region has been indicated by the gray-shaded area in the Figure) as possible proto-GC systems. This condition was chosen in order to have a broad enough assessment of stellar systems with a relatively low  but non-zero DM fraction. It is indeed possible that some GCs formed within an extended DM halo \citep[e.g.][]{Rosenblatt:1988aa}. The latter objects would be still included in our analysis by allowing a baryon fraction somewhat lower than unity. 

We have not chosen to include an $F_{\star}$ cut, because it is expected that most of the gas within these objects will turn into stars at times subsequent to the final redshift of the simulation, eventually turning them into bona fide GCs. The selected objects have a baryonic mass between $1.8 \times 10^{4}$ and $3.7 \times 10^{6}$~M$_{\sun}$, which is still in the mass range of the expected stellar mass of GCs at $z=0$ \citep[e.g.][]{Harris:1991aa, Elmegreen:1997aa, Fall:2001aa, Brodie:2006aa, Mclaughlin:2008aa, Elmegreen:2010aa}.

\subsubsection{Number density}\label{sec:num}

\begin{figure}
\centering
\setlength\tabcolsep{2pt}%
\includegraphics[ trim={0cm 0cm 0cm 0cm}, clip, width=0.48\textwidth, keepaspectratio]{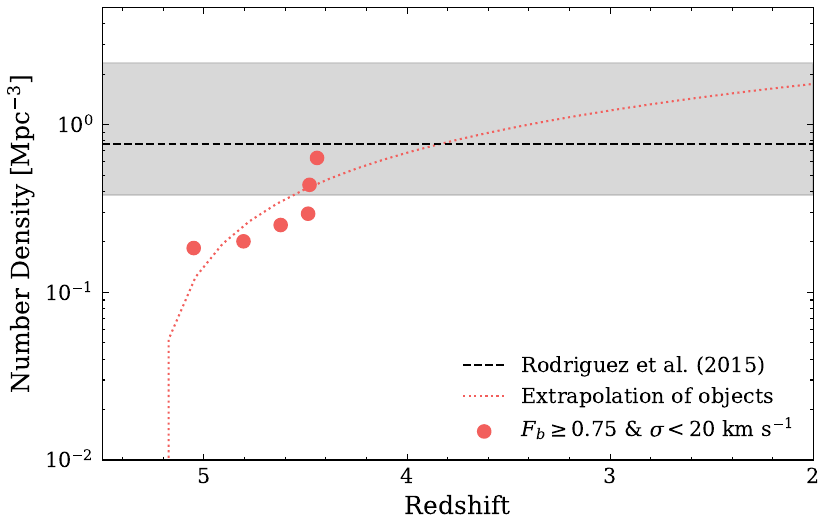}
\caption{The evolution of the number density of potential proto-GCs in the simulation (as explained in Sections~\ref{sec:class} and \ref{sec:prop}). An estimate of the local number density of GCs is indicated by the black dashed line at 0.77~Mpc$^{-3}$, where the gray shaded area shows the error margins presented in \citet{Rodriguez:2015aa}. The dotted red line shows the temporal evolution of number density in our simulation, assuming a linear relation between number density and redshift.}
\label{fig:Numdens}
\end{figure}

In Figure~\ref{fig:Numdens}, we plot the evolution of the number density, $\Phi$, as a function of redshift, for the halos with stellar mass $10^4 \leq M_{\star}[{\rm M}_{\sun}] \leq 10^8$ with $F_{\rm b} \geq 0.75$ and stellar velocity dispersion $\sigma_{\star} < 20$~km~s$^{-1}$ (see Section~\ref{sec:prop}). The number density is computed by dividing the total number of objects ($= 9$) by the volume of the high-resolution region of the simulated box, which is approximately 2 Mpc$^{3}$ at $z=4.4$. An observed estimate of the number density of GCs in the local Universe, within a volume of $\sim 10^{3.6}$~Mpc$^{3}$, is indicated by the black dashed line at 0.77 Mpc$^{-3}$, and the lower and upper error limit are indicated by the gray-shaded area \citep[for more details, see][]{Rodriguez:2015aa}.

\begin{figure*}
\centering
\setlength\tabcolsep{2pt}
\includegraphics[ trim={0cm 0cm 0cm 0cm}, clip, width=1\textwidth, keepaspectratio]{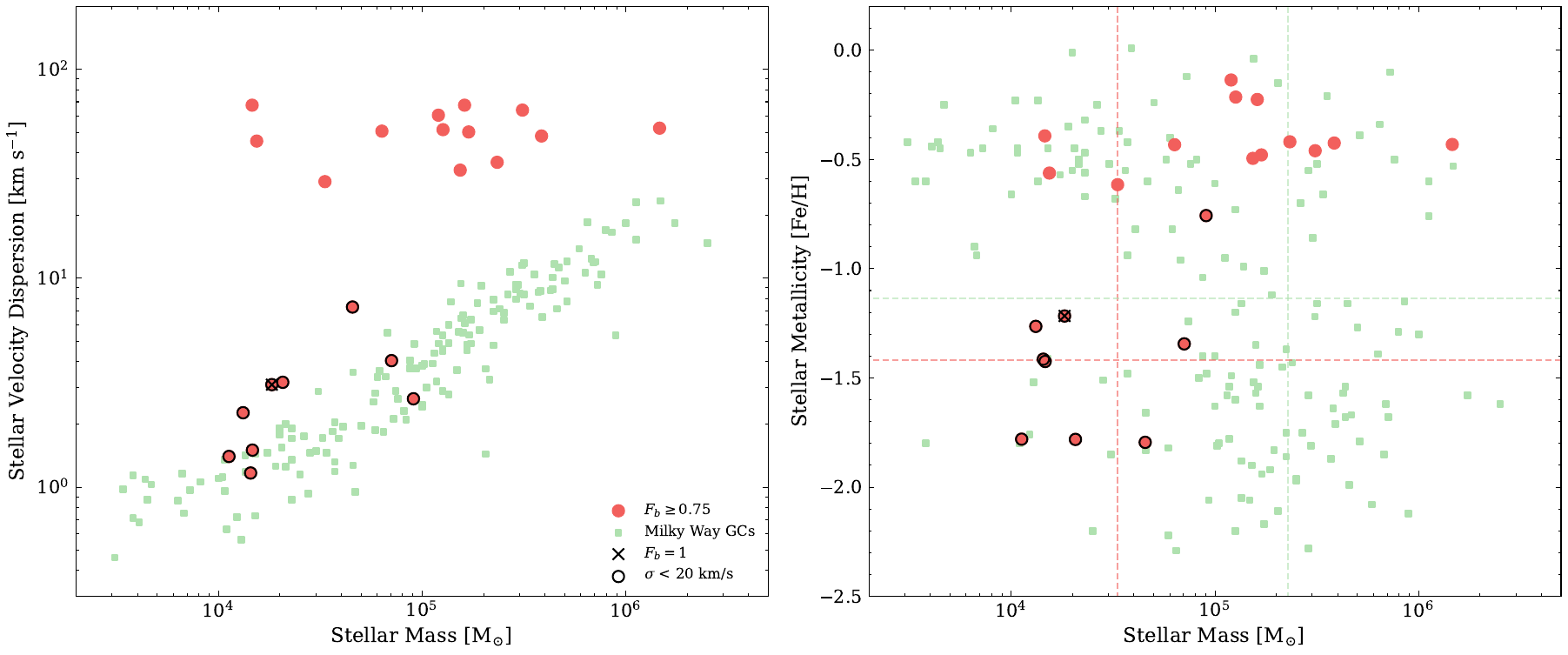}
\caption{Stellar velocity dispersion (left-hand panel) and stellar metallicity (right-hand panel) as a function of stellar mass. The red dots indicate the possible proto-GCs defined by applying the baryonic-mass-fraction selection criterion (see Section~\ref{sec:class}). The red dots with a black outline represent the objects with $F_{\rm b} \geq 0.75$ and $\sigma_{\star} < 20$~km~s$^{-1}$ samples. The black `x' indicates the object with $F_{\rm b} =1$. Corresponding quantities for Galactic GCs are overlaid as green squares taken from \citet{McLaugh:2005aa}. The straight lines in the right-hand panel show the mean stellar mass (vertical) and stellar metallicity (horizontal) of the Galactic GCs (green) and of the $F_{\rm b} \geq 0.75$ and $\sigma_{\star} < 20$~km~s$^{-1}$ objects (red).}
\label{fig:prop}
\end{figure*}

We estimate the number density value to be $\approx 0.63$~Mpc$^{-3}$ at $z=4.4$. By comparing this to the value estimated in the literature, this is slightly lower than what has been approximated by \citet{Rodriguez:2015aa} at present day ($\approx 0.77$~Mpc$^{-3}$), suggesting that there would still be a fraction of GCs that will form after $z = 4.4$. Indeed, GCs are expected to still be able to form till $z \sim 3$ \citep{Katz:2013aa, Kruijssen:2015aa}, which would make our number density at $z=4.4$ a reasonable estimate. By assuming a linear relation between $\Phi$ and $z$, we can approximate the trend between the redshift and number density, finding ${\rm d}\Phi/{\rm d}z = -0.53$~Mpc$^{-3}$. From this we can determine that we would expect $\Phi= 0.77$~Mpc$^{-3}$ at $z=3.8$. Changing the definition of proto-GCs to halos with $F_{\rm b} \geq 0.6$ and stellar velocity dispersion $\sigma_{\star} < 20$~km~s$^{-1}$, we find ${\rm d}\Phi/{\rm d}z = -1.61$~Mpc$^{-3}$ and $\Phi= 0.77$~Mpc$^{-3}$ at $z=5.1$. 

Nevertheless, one has to note that \citet{Rodriguez:2015aa} used a much larger volume than the volume used to calculate the GC number density of GigaEris. The latest  observations have estimated there to be $>150$ GCs within the Milky Way's vicinity \citep[e.g.][]{Baumgardt:2018aa,Vasiliev:2021aa}, from which most can be found within the virial radius of the Milky Way \citep[e.g][]{Zinn:1985, Vasiliev:2021aa}. Hence, one would expect the number density within a 2~Mpc$^{3}$ cube around a Milky Way-like galaxy to be higher than $\approx 0.77$~Mpc$^{-3}$. Therefore, we can conclude that the criteria used to select proto-GCs, bound halos with stellar mass $10^4 \leq M_{\star}[{\rm M}_{\sun}] \leq 10^8$ with $F_{\rm b} \geq 0.75$ and stellar velocity dispersion $\sigma_{\star} < 20$~km~s$^{-1}$, does not overestimate the amount of GCs in a Milky Way-like galaxy. Therefore, we are confident that the simulated proto-GCs in this study are in fact part of the proto-GCs that will become GCs in the local Universe at $z=0$.

\subsection{Properties} \label{sec:prop}

We now study the properties of our proto-GC candidate systems. These properties will allow us to investigate the hypothesis that the $F_{\rm b} \geq 0.75$ objects are indeed possible proto-GCs. Figure~\ref{fig:prop} shows how these objects relate to their stellar mass ($M_{\star}$) in relation to observations of Milky Way GCs. 

In the left-hand panel, the stellar velocity dispersion-mass plane ($\sigma_{\star}$ versus $M_{\star}$) for the $F_{\rm b} \geq 0.75$ objects has been plotted, along with the corresponding observational data for the Milky Way GCs taken from \citet{McLaugh:2005aa}. The velocity dispersion is calculated in comparison to the nearest neighbours for each cluster. As shown by the observational data in Figure~\ref{fig:prop}, GCs in the Milky Way are expected to have a stellar velocity dispersion of $\lesssim$20~km~s$^{-1}$. Although the $\sigma_{\star}$ of the GCs may decrease with time, as GCs lose stellar mass via dynamical ejection of stars through two-body relaxation and the stripping of stars by the galactic field, we exclude the $F_{\rm b} \geq 0.75$ objects with a velocity dispersion higher than 20~km~s$^{-1}$ from our sample. This is because the excluded objects can be found within 1~kpc from the galactic centre of the main galaxy halo. One possibility is that these objects are predecessors of the central galaxy's nuclear star cluster. This possibility is further explored in \citet{vanDonkelaar:2023aa}. The proto-GCs with $\sigma_{\star} < 20$~km~s$^{-1}$ are in good agreement with the observations. As the properties of our proto-GCs are taken at $z=4.4$, it is reasonable that the objects in our sample are at the lower end of the mass range of the present-day observations.

The right-hand panel of Figure~\ref{fig:prop} shows the stellar metallicity against the stellar mass for the extracted  objects and observational data from Galactic GCs. Our whole sample (including the systems with $\sigma_{\star} > 20$~km~s$^{-1}$) appear to be consistent with measurements of metallicities of present-day GCs. The proto-GCs, however, have a mean metallicity lower than what we find in the present-day Milky Way GCs (by $\sim$ 0.6 dex).  This is reasonable, as GCs in the Milky Way have a bimodal metallicity distribution, with the old (and blue) metal-poor population outnumbering the metal-rich ones. The proto-GCs have a similar metallicity as the blue GC population in the Milky Way \citep[e.g.][]{Tonini:2013aa}. Furthermore, as we have no GCs that formed later than $z\sim 4.8$ and GCs are still expected to form till  $z \sim 3$ \citep[][]{Katz:2013aa,Kruijssen:2015aa}, when metallicity is higher.

\begin{figure}
\centering
\setlength\tabcolsep{2pt}%
\includegraphics[ trim={0cm 0cm 0cm 0cm}, clip, width=0.45\textwidth, keepaspectratio]{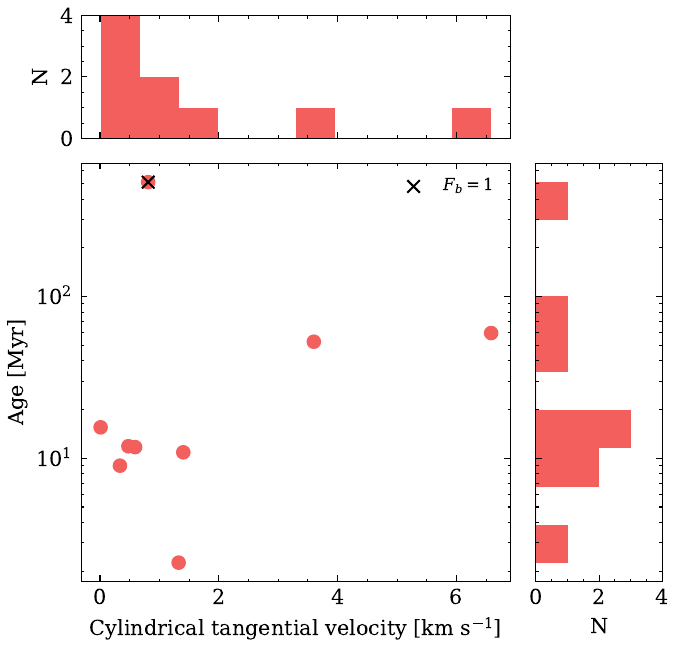}
\caption{Mean stellar age of the proto-GCs against the cylindrical tangential velocity at the outer edge of the clusters. The black `x' indicates the object with $F_{\rm b} =1$. The top and right panels show the full cylindrical tangential velocity and mean stellar age distributions for the proto-GCs with $F_{\rm b} \geq 0.75$ and $\sigma_{\star} < 20$~km~s$^{-1}$.}
\label{fig:rot}
\end{figure}

In Figure~\ref{fig:rot}, we present the mean age of stars in the proto-GCs at $z=4.4$ against the cylindrical tangential velocity of the most outer stars of the cluster together with the distributions of the two properties. It is expected that GCs are in general old \citep[e.g.][]{Trenti:2015aa} and have low cylindrical tangential velocities \citep[e.g.][]{Bianchini:2013aa}.  There is no clear relation between the two properties. The figure shows that the $F_{\rm b} = 1$ object is much older than the other objects, with a mean age of 508~Myr, whereas the other objects have a mean age between 2 and 59~Myr. Furthermore, the cylindrical tangential velocity is low for all objects, with most objects having $V_{\rm t} < 2$~km~s$^{-1}$ and two outliers with 3.6 and 6.6~km~s$^{-1}$. As GCs are formed at high redshift, we would expect that any rotation signature would have been eradicated by $z=0$. In the recent years, previous measurements \citep[e.g.][]{Lane:2011aa, Bellazzini:2012aa} have shown some rotation in a couple of systems by probing only the motion of the stars in the outer edges. Therefore, the determined cylindrical tangential velocity is compatible with observations of present-day GCs. The full list of the different properties explored for each object can be found in Table~\ref{tab:prop}.

\begin{figure}
     \centering
     \begin{subfigure}[b]{0.47\textwidth}
         \centering
         \includegraphics[width=\textwidth]{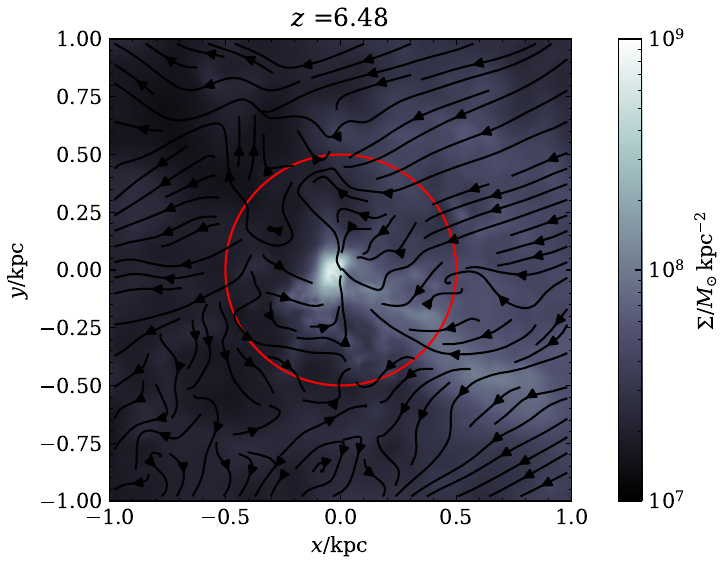}
         \caption{object 6}
         \label{fig:halo6}
     \end{subfigure}
     \hfill
     \begin{subfigure}[b]{0.47\textwidth}
         \centering
         \includegraphics[width=\textwidth]{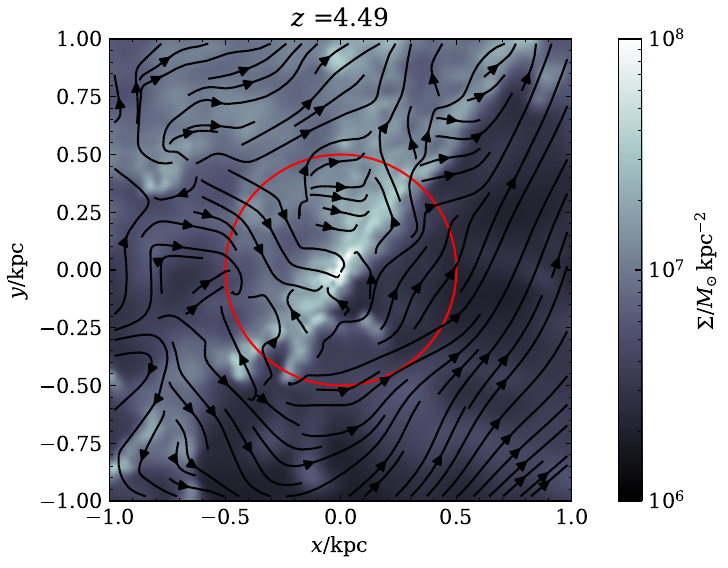}
         \caption{object 2}
         \label{fig:halo2}
     \end{subfigure}
     \hfill
     \begin{subfigure}[b]{0.47\textwidth}
         \centering
         \includegraphics[width=\textwidth]{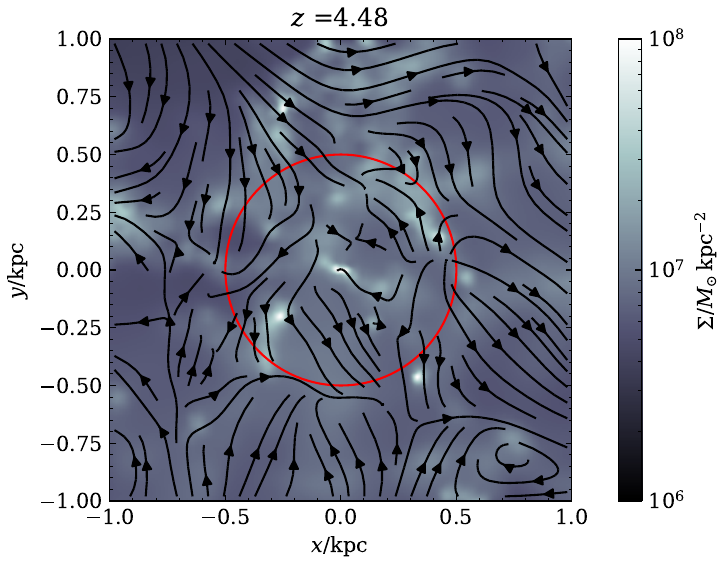}
         \caption{object 5}
         \label{fig:halo5}
     \end{subfigure}
        \caption{Gas surface density maps one time step before the formation of the first stars in the proto-GC, with the velocity vectors of the gas overlaid. From top to bottom, the objects projected are object~6 (\ref{fig:halo6}), object~2 (\ref{fig:halo2}), and object~5 (\ref{fig:halo5}). The gas map is centred on the centre of mass of the gas particles that are predecessors of the stars in the proto-GC at $z=4.4$. The red circle indicates a sphere with a radius of 0.5~kpc.}
        \label{fig:gastreams}
\end{figure}

\subsection{Formation and evolution}\label{sec:form}

\begin{figure*}
\centering
\setlength\tabcolsep{2pt}
\includegraphics[ trim={0cm 0cm 0cm 0cm}, clip, width=1\textwidth, keepaspectratio]{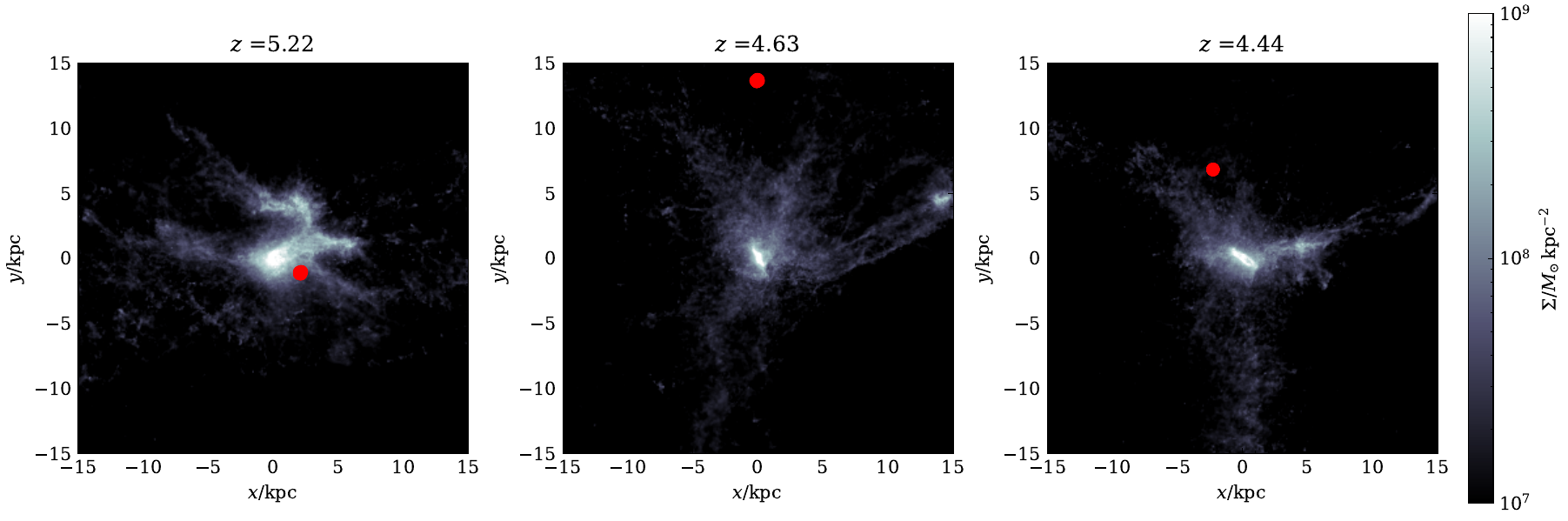}
\caption{Gas surface density maps of the main galaxy halo, with object~6  indicated in red, at different redshifts. At $z=5.22$ (left-hand panel), object~6  is at its closest point from the centre of the main galaxy halo. From $z=4.63$ to $4.44$ (middle and right-hand panels), the object is again moving towards the centre of the main galaxy halo, it is during this period that most of its DM content is being stripped (see Figure~\ref{fig:DMevolution}).} 
\label{fig:evolution}
\end{figure*}

\begin{figure}
\centering
\setlength\tabcolsep{2pt}%
\includegraphics[ trim={0cm 0cm 0cm 0cm}, clip, width=0.45\textwidth, keepaspectratio]{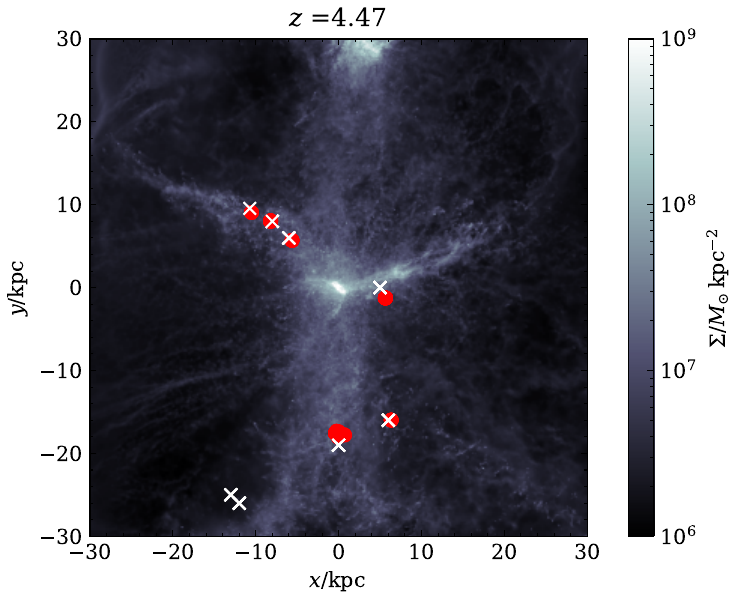}
\caption{Gas surface density map centred on the main galaxy halo at $z = 4.47$. The red dots indicate the location of the proto-GCs, excluding object~6 , at $z = 4.47$. The white crosses indicate their birth location. The  three dots at (0,-18)~kpc correspond to the crosses at (0,-18)~kpc and (-13,-27)~kpc.}
\label{fig:birth}
\end{figure}

To investigate the formation of the proto-GCs, we analyze the gas surface density\footnote{The depth of the surface density plots is the depth of the simulated box.} of the environment where the stellar structures formed. The surface density is of interest as the local SFR per unit area is linearly proportional to the local surface density of dense gas \citep[][]{Dobbs:2009aa}. Figure~\ref{fig:gastreams} shows the gas surface density just before the formation of the first stars, with the velocity vectors of the gas on top. The red circle indicates the 0.5~kpc sphere in which the formation of the first stars will happen. The top panel in Figure~\ref{fig:gastreams} shows object~6 (Figure~\ref{fig:halo6}), which is the object with $F_{\rm b} = 1$. The gas properties of the formation environment of this object are different from the other two objects (Figure~\ref{fig:halo2}: object~2; Figure~\ref{fig:halo5}: object~5). The main gas halo of Figure~\ref{fig:halo6} has a higher gas mass and seems to be more isolated in comparison to the other two objects in Figures~\ref{fig:halo2} and \ref{fig:halo5}. 

The formation structure of object~6 also seems different from the other two. From Figure~\ref{fig:halo6}, we can identify that object~6 has one large gas tail in the bottom-right corner, from which the part that is within the 0.5~kpc radius is inflowing. This inflowing gas has a temperature of $\sim$10$^4$~K. On the other hand, object~2 has a high density gas filament of mostly cold outflowing gas ($\sim 10^2$ K) towards the upper right corner. object~5 is surrounded by many other small gas clouds. The inflowing gas of object~2 and object~5 have a similar temperature as that of object~6 , $\sim$10$^4$~K. We find that all proto-GCs are found within 5 kpc away from a DM subhalos, which could be in line with the formation mechanism discussed by \citet{Madau:2020aa}. Especially as the proto-GCs in this work show similar stellar masses as found by \citeauthor{Madau:2020aa}. However, further research is needed to determine if these DM subhalos did influence the formation of the proto-GCs, by tracking the orbits of these DM halos and determining if they collided.

As expected for GCs, the SF history of the objects shows one big burst of SF. For all objects, this happens within a time frame of $\sim$ 15 Myr with a specific SFR between $0.05$ and $1.05$ Myr$^{-1}$, with an average of $0.42$ Myr$^{-1}$ for all objects. This means that the proto-GCs within this study have a single stellar population at $z=4.44$ and therefore will most likely not show any multi-model metallicity distribution at $z=0$. Additionally, all the proto-GCs, except object~6, are still forming stars at $z = 4.4$. Object~6 forms stars for $\sim$10~Myr, starting at $z \approx 6.5$. 

Zooming out, we can see that between redshift $z\approx 4.80$ and $z\approx 4.44$ the main galaxy halo experiences a minor merger (see the middle and right-hand panel of Figure~\ref{fig:evolution}). Tantalisingly, all proto-GCs, excluding object~6, were born between $z=4.64$ and $z=4.47$. We therefore theorize that the gas inflow into the system as a result of the merger could have had an influence on the formation of these objects. Figure~\ref{fig:birth} displays the birth location of the objects, excluding object~6, together with their location at $z=4.47$. From this we can conclude that the spatial distribution of the objects resembles a filamentary network of multiphase gas associated with cooling of the intracluster gas \citep[see also][]{Lim:2020aa} and thus the proto-GCs formed within the gas filaments. During this timeframe we also see a slight increase in the fraction of cold ($T<$10$^5$~K) gas within a 80 kpc box surrounding the main galaxy halo. This would also be in line with the GC formation scenario wherein GCs are formed as a byproduct of active SF in the galaxy  \citep[e.g.][]{Elmegreen:2010aa, Shapiro:2010aa, Kruijssen:2015aa}, when gas is compressed.

The objects formed all between 10 to 30~kpc away from the galactic centre of the main halo. The difference between their starting and ending (at $z = 4.4$) radius away from the centre is on average 3~kpc. However, this is significantly increased by objects~3, 6, and 8, which all covered a minimum distance of $\sim$11~kpc. 

\subsubsection{The imposter}

From the formation history discussed in Section~\ref{sec:form}, we can conclude that object~6  is an outlier as it formed at a higher redshift, it was more isolated, and had a different formation structure. For all the other properties discussed, like stellar mass, stellar velocity dispersion, metallicity, and rotation, the object is similar to the rest of the proto-GCs. The only hints we have towards the different nature of this object are its high mean age and different formation. Additionally, next to $F_{\rm b} = 1$ for this object, $F_{\star}$ is also 1 at $z=4.4$.

\citet{Phipps:2020aa} saw a similar object within their simulations and speculated it could be a forming bulge or nuclear star cluster of the main galaxy. However, our object formed 15~kpc away from the centre of the main galaxy halo, so this is unlikely. Furthermore, from Figure~\ref{fig:evolution}, we can see that this structure has interacted with the main galaxy halo. We therefore speculate it to be some type of bulge outside the main galaxy halo, which due to gravitational interaction has been stripped of all its DM {\it and} gas. 

\begin{figure}
\centering
\setlength\tabcolsep{2pt}%
\includegraphics[ trim={0cm 0cm 0cm 0cm}, clip, width=0.45\textwidth, keepaspectratio]{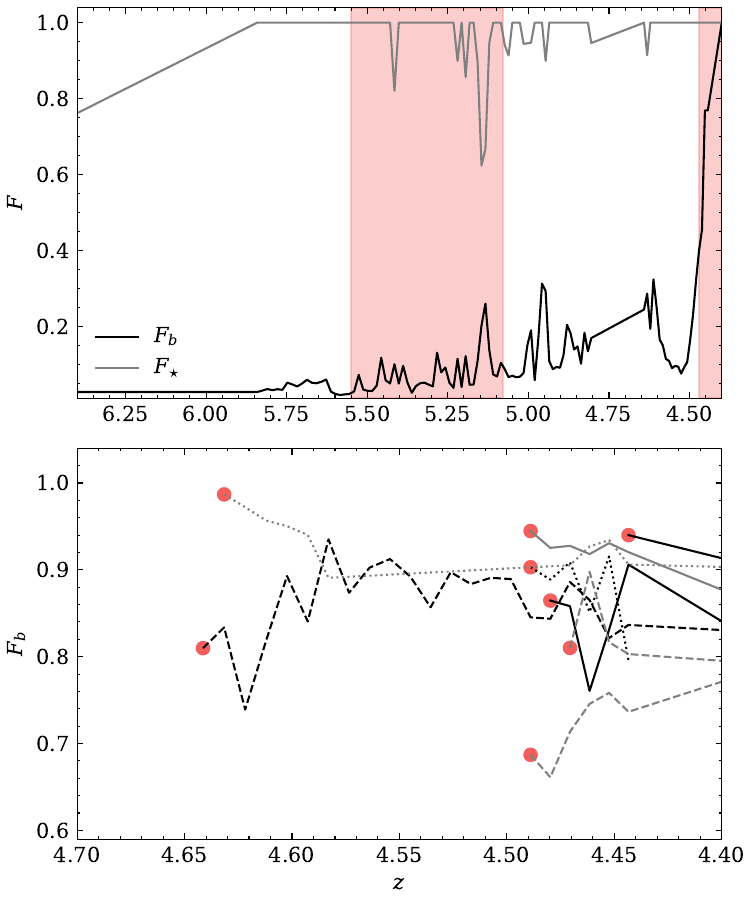}
\caption{ Mass fraction versus redshift. The top panel shows the evolution of the baryonic (black line) and stellar (gray) mass fraction for object~6 . The red shaded area indicates when the object is within 10 kpc of the centre of the main galaxy halo (see also Figure~\ref{fig:evolution}).  The bottom panel shows the baryonic mass fraction evolution for  objects~1--5 and objects~7--9. The red dot indicates the baryonic mass fraction of the system when it was born.}
\label{fig:DMevolution}
\end{figure}

This is supported by the $F_{\rm b}$ and $F_{\star}$ evolution of this object, as shown in the top panel of Figure~\ref{fig:DMevolution}. The top panel shows this evolution for object~6 , the red shaded area indicates when the object is gravitationally interacting with the main halo, which we define as the object being within 10 kpc of the centre of the main galaxy halo (see also Figure~\ref{fig:evolution}). The moment the object enters this region, the DM starts to be stripped away from the stellar cluster. There is a very steep increase in baryonic mass fraction from $z \sim 4.6$ for object~6 . During this period, the cluster falls back into the main galaxy halo, with the distance to the galactic centre decreasing from 14 to 7.3~kpc. This last event had as a result that the rest of the DM has been stripped from object~6 .

This kind of DM evolution is unique for object~6  as shown by the baryonic mass fraction evolution of the other objects, plotted in the bottom panel of Figure~\ref{fig:DMevolution}. In comparison to object~6 , these objects are already born with a low DM fraction and stay at a similar baryonic mass fraction during their evolution. Therefore, we can conclude that object~6  went through a different DM evolution in contrast to the other objects, and thus probably is a different astrophysical object. 

One possibility is that object~6  is an ultra compact dwarf galaxy (UCD) that has been tidally stripped of its DM during the interaction with the main galaxy halo. As UCDs are stellar systems that bridge the divide between  GCs and dwarf galaxies \citep[e.g.][]{Drinkwater:2003aa}, this would explain the difference in age, formation, evolution, and stellar fraction with the other proto-GCs. UCDs are systems with effective radii of $10 < R_{\rm eff}/{\rm pc} < 100$ \citep[][]{Brodie:2011aa}, which is consistent with the radius of object~6 , which is $28$ pc.  UCDs exhibit high dynamical masses for objects of their size, $\sim$ $10^6$--$10^8$~M$_{\sun}$ \citep[][]{Mieske:2008aa}. object~6  is a lot less massive, with a total mass of $1.8 \times 10^4$~M$_{\sun}$. However, this could be the result of the stripped DM. Another possibility is that object~6  was a nucleus of a dwarf satellite galaxy that was tidally stripped of gas and DM by its host galaxy.  Currently, one observed Galactic GC is being theorized to be the nucleus of an accreted galaxy, $\omega$ Centauri \citep[][]{Freeman:1993aa, Bekki:2003aa}. Nevertheless, object~6  is very different from $\omega$ Centauri. For example, object~6  does not show any sign of rotation or multi-model metallicity distribution which is instead observed in $\omega$ Centauri \citep[e.g.][]{Norris:1995aa, 3Ms:1997aa}. Therefore, we speculate that object~6  and $\omega$ Centauri have different formation histories.

A property not yet explored in this study is the fraction of ``born thin-disc'' stars\footnote{To determine if a star was born in the thin disc, we use the stellar IDs of stars  identified by \citet{Tamfal:2022aa} using the \textsc{DBSCAN} \citep{Ester:1996aa} clustering algorithm.} in each object. Tantalisingly, object~6  is the only identified object with $F_{\rm b} \geq 0.75$ and $\sigma_{\star} < 20$~km~s$^{-1}$ that has a non-zero percentage of these stars (17 per cent). This fraction of stars were born in the thin disc at $z \sim 6$, before the first interaction of the object with the main galaxy halo (at $z \sim 5.3$). Whereas some scenarios predict a formation with the geometric thin disc significantly forming after the thick disc has formed \citep[e.g.][]{Bird:2013aa}, more recent studies find that the thin disc is expected to start forming at the earliest assembly stage as shown in \citeauthor{Tamfal:2022aa} (\citeyear{Tamfal:2022aa}; see also \citealt{Agertz2021, Silva:2021aa, Michael:2022aa,  vanDonkelaar:2022aa} for the early formation of a thin-disc component). Therefore, having thin-disc stars form at $z \sim 6$ is possible and from this we can conclude that these stars formed in the main galaxy halo. During the interaction between object~6  and the main halo, the stars got then trapped by the gravitational potential of object~6  and left the main galaxy halo. This has as a result that not all the stars within this object have been born in the same environment and that object~6  had an influence on the main galaxy halo.

\section{Discussion} \label{sec:disc}

We explored the formation, evolution, and properties of proto-GCs at $z>4$ through the use of the cosmological, $N$-body hydrodynamical ``zoom-in'' simulation GigaEris \citep[][]{Tamfal:2022aa}.  However,  there is no guarantee that all the systems identified in this paper will survive to the present day, as the GCs might get tidally disrupted or merge with each other or the galaxy halo \citep{Forbes:2018aa}. \citet{Gnedin:1997aa}  describe the dominant effects of GC destruction in the Milky Way to be relaxation and tidal shocks through disc and bulge passages. Therefore, at small Galactocentric distances, only massive, compact GCs can survive for more than a Hubble time. The proto-GCs in this work are not very compact as a result of the radii of the proto-GCs, which are at the larger end of what is expected from galactic GCs. This is probably a result of the gravitational softening of the simulation, $\epsilon = 0.043$~kpc at $z=4.4$, which makes the gravity potential of the proto-GCs weaker than it should be. This if further demonstrated by the fact that the crossing time of the proto-GCs is 2.5 times bigger than the free fall time. We therefore assume these radii to be an upper range. Furthermore, as the objects are still star-forming at $z=4.4$, we expect the objects to get more dense over time. This would make it harder for the GCs to be  tidally disrupted by gravitational pull of the main galaxy. Additionally, the proto-GCs in this study form at locations that would likely allow them to survive to the present day, as they are at large enough distance from the centre of the main galaxy halo.  Lastly, the dynamical friction time scale of these objects, $\sim 10^{14}$~yr, is larger than Hubble time. This all shows us that the proto-GCs will most likely all surive to the present day.  

Although we are unable to follow the evolution of the proto-GCs to $z=0$ and therefore cannot make definitive statements about whether these objects are truly GC analogs, many of the discussed properties are grossly consistent with the present-day GCs. One clear difference from observations is the metallicity of the objects in our simulations, which on average is slightly lower than the average metallicity of the GCs observed in the Milky Way \citep[][see Figure~\ref{fig:prop}]{McLaugh:2005aa}. However, most of our candidates are in a phase of their evolution when their SF has not yet fully finished, thus allowing for further evolution of their stellar mass and metallicity. This entails that, for example, our proto-GCs could still move on the stellar metallicity-mass plane and hence could eventually better resemble the present-day GC observations of the Milky Way. Furthermore, the proto-GCs have a similar metallicity to that of the blue GC population in the Milky Way \citep[e.g.][]{Tonini:2013aa}, meaning we are missing the red GCs within our sample. This is supported by the seemingly low number density within the simulated box at $z=4.4$. None of the detected proto-GCs formed within the disc of the main galaxy halo, implying that we are missing the GCs formed within giant molecular clouds \citep[GMCs; see, e.g.][for the modelling of the formation of GCs in GMCs]{Li:2017aa, Grudic:2021aa, Grudi:2023aa}. Additionally, one would expect that a fraction of the GCs  will form after $z = 4.4$, as GCs are predicted to form till $z \sim 3$ \citep{Katz:2013aa, Kruijssen:2015aa}. Lastly, we also assume that with a higher resolution one would detect more proto-GCs within the simulation. 

While our work covers a different redshift and simulation resolution from what has previously been studied in full cosmological simulations that resolve the formation of proto-GCs \citep[e.g.][]{Kim:2018aa, Ma:2020aa, Phipps:2020aa, Omid:2022aa}, our results are broadly consistent with those from these previous numerical simulations. For example, \citet{Phipps:2020aa} found a number density of $\sim$0.82~Mpc$^{-3}$ at redshift 4.4 for objects with $F_{\rm b} \geq 0.80$, which is similar to what we  expect ($\Phi = 0.63$ Mpc$^{-3}$ for objects with $F_{\rm b} \geq 0.75$).

Furthermore, in our simulation, the proto-GC gas clouds formed through inflowing gas and the stars then formed in a region with a gas density of $\sim$ 10$^8$~M$_{\sun}$~kpc$^{-2}$, as shown by Figure~\ref{fig:gastreams}. This is consistent with \citet{Omid:2022aa}, who found that cluster formation is triggered when dense, turbulent gas reaches $\Sigma_{\rm gas} \approx 10^7$ M$_{\sun}$~kpc$^{-2}$ at $z>4$. \citet{Ma:2020aa} also found this density threshold for galaxies at $z\geq 5$, using high-resolution cosmological ``zoom-in'' simulations from the FIRE project. Additionally, \citeauthor{Ma:2020aa} show that these high-pressure clouds are compressed by collision of gas streams in highly gas-rich turbulent environments, something we have also seen for the formation of proto-GCs at $z>4$. 

\subsection{Detectability at high redshift}

The identified proto-GCs in this study are most likely not massive enough to be detected by JWST or ALMA at $z>4$, as we probably need stellar clusters with a stellar mass of minimum  $\sim10^6$~M$_{\sun}$ \citep[for JWST, as shown by][]{Vanzella:2022aa}. A way to access this low-mass regime is the exploitation of gravitational lenses. Strong lensing amplification will allow us to probe the structural parameters down to the scale of a few tens of parsec \citep[e.g.][]{Livermore:2015aa,Rigby:2017aa, Vanzella:2017aa, Vanzella:2019aa} and observe clustered star-forming regions that are  otherwise not spatially resolved. With strong lensing, we will need stellar clusters with a stellar mass of minimum  $\sim10^5$~M$_{\sun}$ to be able to observe them \citep[as shown by][]{Calura:2021aa, Vanzella:2022ab}, which means that some of our high-mass proto-GCs could possibly be detected. Furthemore, if a similar SFR continues for each proto-GC, they will have a stellar mass of $\sim10^6$~M$_{\sun}$ at $z\sim3$. This would significantly increase their chance of being detected at $z>3$.

The total baryonic mass of the identified objects, exluding object~6, is between $5 \times 10^{5}$ and $3 \times 10^{6}$~M$_{\sun}$. Therefore, even if the proto-GCs in this study would transform all the gas into stars, observing them at high redshift would still be difficult. However, there is a possibility that the higher-mass objects can be observed with ALMA through the detection of strong far-infrared lines (e.g. the [CII] line, useful to trace star-forming gas; \citealt{DeLooze:2014aa, Calura:2021aa}), when the GCs are massive enough and still star forming. 

\subsubsection{Identifying the imposter, and implications on the GCs-dwarf galaxies connection}
Important to note is that with current observation techniques we would not have been able to identify object~6  as a different object than a GC at $z=0$. This is because its measurable properties perfectly overlap with the observed GCs in the current Milky Way, as shown in Figure~\ref{fig:prop}. Therefore, there is a significant chance that we have objects like object~6  in our current Galactic GC catalogue. The detectability of such a cluster with the recently launched JWST at high redshift should be explored in the future. \citet{Pozzetti:2019aa} have shown that the possibility to detect the precursors of GCs at their actual formation redshift is small. However, we have speculated object~6  to be a UCD or nucleus of a dwarf galaxy that has been stripped of its gas and DM: this could increase the likelihood of detection at higher redshift. 

In order to investigate if there are more ``imposter''-like objects in the simulation, we studied the evolution of the DM content of other $F_{\star}=1$  bound systems identified with AMIGA . These systems show a completely different evolution, forming with a a higher $F_{\rm b} \sim 0.2$ to $\sim 0.7$, and follow a similar evolution to that exhibited by the proto-GCs in the simulation. Namely, they maintain a similar baryonic mass fraction as when they were formed.  Unlike object~6 , the other $F_{\star}=1$ objects remain at a similar distance from the main galaxy halo from birth to $z=4.4$. As a result, they do not undergo significant tidal heating, retaining their DM. These systems are likely associated with the population of faint dwarf galaxy satellites, including the ultra faint dwarfs (UFDs), and their properties will be studied in a subsequent paper. From this analysis, we can conclude that ``the imposter'' is a unique object in the simulation, although we cannot exclude that similar systems might arise in the later part of the evolution  which is not yet covered by our simulation.  Placing these results in the general context, they show how the separation between GCs and dwarf galaxy satellites can be rather loose when one considers their formation histories, consistent with the observational fact that there is a significant degree of continuity in terms of structural properties among different types of stellar systems that are classified as either GCs or as one of the several  morphological types of dwarf galaxies.

\section{Conclusions} \label{sec:conc}

Using a high-resolution cosmological ``zoom-in'' simulation of a Milky Way-sized galaxy halo \citep[GigaEris;][]{Tamfal:2022aa}, we study the properties of possible proto-GCs. Our main conclusions are as follows.

\begin{itemize}

    \item We define proto-GC systems as objects with $F_{\rm b} \geq 0.75$ and $\sigma_{\star} < 20$~km~s$^{-1}$. They have a relatively low stellar mass ($M_{\star} \sim 10^4$--$10^5$~M$_{\sun}$) in comparison to present-day Galactic GCs and a metallicity ($-1.8 \lesssim {\rm [Fe/H]} \lesssim -0.8$) similar to the blue Galactic GCs \citep[e.g.][]{Tonini:2013aa}. They show minimal rotation, with most identified clusters having $V_{\rm t} < 2$~km~s$^{-1}$. Lastly, all objects but object~6  still have some DM in their halo at $z=4.4$.
    
    \item The number density of proto-GCs at $z=4.4$ is approximately 0.63~Mpc$^{-3}$. This is lower than what has been approximated by \citet{Rodriguez:2015aa} at $z=0$ ($\approx$0.77~Mpc$^{-3}$) in the local Universe. This is in agreement with observations in the current Milky Way, as it is known that GCs are still able to form till $z \sim 3$ \citep{Katz:2013aa, Kruijssen:2015aa} and shows that the selection criteria do not overestimate the amount of proto-GCs one could expect. This is supported by the fact that we do not have any proto-GCs with properties similar to those of red GCs.
    
    \item The proto-GC predecessor gas clouds formed in accreting gas filaments in the Circumgalactic region. The stars formed where the gas density exceeded $\sim$ 10$^8$ M$_{\sun}$~kpc$^{-2}$, as shown by Figure~\ref{fig:gastreams}. The objects that formed at $z \leq 5$ are still star-forming at $z=4.4$. All clusters show one big burst of SF and formed between 10 and 30~kpc away from the galactic centre of the main halo. The difference between their starting  and ending radius away from the centre is on average 3~kpc, thus showing minimum movement after birth.

    \item All proto-GCs, excluding object~6, were born between $z= 4.6$ and $z = 4.47$,  during the time when the main galaxy suffers a minor merger. We speculate that increased gas accretion and tidal perturbations excited by the interaction might have played a role in triggering the density enhancements in the filaments that resulted in proto-GC formation.
    
    \item From its formation and evolution, we can speculate object~6  to be a UCD or nucleus of a dwarf satellite galaxy that was tidally stripped of DM by its host galaxy. The stellar cluster was within 10~kpc of the main halo at $5.8 \geq z \geq 5.2$ and $z \geq 4.6$, with the result that the object was tidally stripped of its DM and attracted stars born in the thin disc of the main galaxy halo in its gravitational potential.
    
    \item With current observation techniques, we would not have been able to identify object~6  as a different astrophysical object than a GC. This is because its measurable properties at $z=0$ perfectly overlap with the observed GCs in the current Milky Way. 
    
\end{itemize}

\section*{Acknowledgements}
We thank the anonymous Reviewer for their constructive input. PRC, LM, and FvD acknowledge support from the Swiss National Science Foundation under the Grant 200020\_192092. We acknowledge Eros Vanzella for insightful discussions. 

%%%%%%%%%%%%%%%%%%%%%%%%%%%%%%%%%%%%%%%%%%%%%%%%%%
\section*{Data Availability}
The data underlying this article will be shared on reasonable request to the corresponding author.



\bibliographystyle{mnras}
\bibliography{paper} % if your bibtex file is called example.bib

%\begin{thebibliography}{99}
%\end{thebibliography}
%%%%%%%%%%%%%%%%%%%%%%%%%%%%%%%%%%%%%%%%%%%%%%%%%%

%%%%%%%%%%%%%%%%% APPENDICES %%%%%%%%%%%%%%%%%%%%%

\appendix

\section{Properties of identified objects}
Table~\ref{tab:prop} summarizes the properties of the halos with $F_{\rm b} \geq 0.75$ and stellar velocity dispersion $\sigma_{\star} < 20$~km~s$^{-1}$ identified by AHF.

\begin{table}
\centering
\begin{tabular}{cccccccc}
\hline
Object & $F_{\rm b}$ & $F_{\star}$ & $\log_{10}(M_{\star}$) & $R_{\rm m}$  & $\sigma_{\star}$ & $Z_{\star}$  & $V_{\rm t}$ \T \B \\
&  &  & [${\rm M}_{\sun}$] & [pc] & [km~s$^{-1}$] & [Fe/H] & [km~s$^{-1}$] \T \B \\\hline
1 & 0.91 & 0.01 & 4.05 & 16  & 1.40 & -1.78 & 1.32 \T \B \\
2 & 0.88 & 0.06 & 4.96 & 114 & 2.65 & -0.76 & 0.59 \T \B \\
3 & 0.83 & 0.08 & 4.31 & 428 & 3.18 & -1.78 & 6.58 \T \B \\
4 & 0.77 & 0.01 & 4.12 & 132 & 2.27 & -1.26 & 0.01 \T \B \\
5 & 0.85 & 0.03 & 4.16 & 32  & 1.17 & -1.41 & 0.48 \T \B \\
6 & 1.00 & 1.00 & 4.26 & 28  & 3.10 & -1.22 & 0.81 \T \B \\
7 & 0.80 & 0.03 & 4.17 & 22  & 1.50 & -1.42 & 0.30 \T \B \\
8 & 0.90 & 0.02 & 4.66 & 252 & 7.28 & -1.80 & 3.60 \T \B \\
9 & 0.79 & 0.11 & 4.85 & 148 & 4.03 & -1.34 & 1.40 \T \B  
\end{tabular}
\caption{Properties of the different objects explored in this paper, measured at $z=4.4$. Column~1: name; Column~2: baryonic mass fraction; Column~3: stellar mass fraction; Column~4: total stellar mass computed at half the virial radius; Column~5: half-mass radius; Column~6: stellar velocity dispersion; Column~7: stellar metallicity; Column~8: cylindrical tangential velocity of the cluster's outer edge.}
\label{tab:prop}
\end{table}

%%%%%%%%%%%%%%%%%%%%%%%%%%%%%%%%%%%%%%%%%%%%%%%%%%

% Don't change these lines
\bsp	% typesetting comment
\label{lastpage}
\end{document}